\documentclass{article}
\usepackage{epsf}
\usepackage{emulateapj}
\begin{document}
\lefthead{Robertson \& Leiter}
\righthead{BHC Magnetic Moments?}

\title{Do Black Hole Candidates Have Magnetic Moments Instead of Event Horizons?}

\author{Stanley L. Robertson\altaffilmark{1} and Darryl J. Leiter\altaffilmark{2}}
\altaffiltext{1}{Dept. of Physics, Southwestern Oklahoma State University,
Weatherford, OK 73096}
\altaffiltext{2}{FSTC, Charlottesville, VA 22901}

\begin{abstract}
In previous work we found that many of the spectral properties of low mass
x-ray binaries (LMXB), including galactic black hole candiates (GBHC)
were consistent with the existence of intrinsically
magnetized central objects. We review and extend these findings and
show that the existence of intrisically
magnetic BHC is consistent with
a new class of solutions of the Einstein field equations of General Relativity.
These solutions are based on a strict adherence to the Strong Principle of
Equivalence (SPOE) requirement that the world lines of physical matter must
remain timelike in all regions of spacetime. The new solutions emerge when the
structure and radiation transfer properties of the energy momentum tensor
on the right hand side of the Einstein
field equations are appropriately chosen to dynamically 
enforce the SPOE requirement of timelike world line
completeness. In this context, we
find that the Einstein field equations allow the existence of highly
red shifted, Magnetospheric, Eternally Collapsing Objects (MECO).
MECO can possess intrinsic magnetic moments since they do not have
trapped surfaces that lead to event horizons and curvature singularities.
Since MECO lifetimes are orders of magnitude greater than a Hubble time,
they provide an elegant and unified framework for understanding a broad
range of observations of GBHC and active galactic nuclei.
We examine their properties and discuss characteristics that might lead to
their confirmation.
\end{abstract}


\section{Introduction}
In earlier work (Robertson \& Leiter 2002) we presented evidence
for the existence of intrinsic magnetic moments of $\sim 10^{29}$ G cm$^3$
in galactic black hole candidates (GBHC). These findings are recapitulated
and extended in Table 1 and Appendix D, which summarizes the equations
and the analysis of observations. Combined with rotation rates in the
range 10 - 50 Hz, these magnetic moments provide a robust unified mechanism
for the spectral state switches observed in GBHC and neutron stars (NS),
a common origin of quiescent power-law emissions as spin-down luminosity, and
a unified driving mechanism for the ubiquitous low-state jets and synchrotron
emissions of both. Magnetosphere topology also serves to stabilize the inner
accretion disk (Arons et al. 1984). The great strength of
the magnetospheric model that we describe is that it
allows a unified description of all of the various spectral and luminosity
states of x-ray novae, whether NS or GBHC. Similarities of NS and GBHC
properties, particularly in low and quiescent states, have been previously
noted, (e.g. van der Klis 1994, Tanaka \& Shibazaki 1996) as well as
the lack of compelling evidence for event horizons (Abramowicz, Kluzniak
\& Lasota 2002). The central
question at this point is, what is the nature of the compact objects that
we call GBHC.

There is
a plethora of piece meal models of these various phenomena. For example,
comptonizing coronae near event horizons, bulk flow comptonization and
magnetic flares on accretion disks have all been invoked to explain the
hard spectral tail of low state GBHC. But the observed ingress/egress
times for dipping sources imply large radiating regions (Church 2001)
that are inconsistent with the compact corona models and can be
consistent with bulk comptonization models only for large scale outflows.
Similarly, radiatively inefficient advective flows at high accretion rates
have been proposed to explain the quiescent power-law emissions of
GBHC, (Narayan et al. 1997, Garcia et al. 2001).
while ignoring the fact that the similar emissions of
accreting millisecond pulsars are adequately explained by magnetospheric
spin-down. Stated more bluntly, Cir X-1, a burster with a magnetic moment
similar to those found for GBHC (Table 1 and Iaria et al. 2001),
exhibits essentially all of the x-ray
spectral and timing characteristics that have been proposed at
various times as distinguishing features of black holes.
A unified model of GBHC
and NS that can be used to clarify subtle differences is clearly needed.

Others have reported evidence for strong magnetic fields in GBHC.
A field in excess of $10^8$ G has been found at the
base of the jets of GRS 1915+105 (Gliozzi, Bodo \& Ghisellini 1999,
Vadawale, Rao \& Chakrabarti 2001). A recent study of optical polarization
of Cygnus X-1 in its low state (Gnedin et al. 2003)
has found a slow GBHC spin and a magnetic field of $\sim 10^8$
gauss at the location of its optical
emission. Given the $r^{-3}$ dependence of field strength on magnetic
moment, the implied magnetic moments
are in good agreement with those we have found. A recent
correlation (Mauche et al. 2002, Warner \& Woudt 2003)
of quasi-periodic oscillation (QPO) frequencies
extending over six orders of magnitude in frequency, from dwarf
novae to neutron stars shows points for GBHC squarely in the middle
of the line. If the higher of the correlated frequencies is generated
where the inner radius of an accretion disk interacts
with a magnetosphere (Goodson, Bohm \& Winglee 1999, Titarchuk \& Wood 2002),
this would be additional evidence of intrinsic magnetic moments for GBHC.
A relativistic frame-dragging explanation is surely not applicable.

While accommodating intrinsic magnetic moments in models of GBHC would
require revisions in the current theoretical
picture of these compact objects,
such an approach would also greatly simplify the problem of understanding the
spectral, timing and jet ejection mechanisms of compact objects.
If GBHC have intrinsic magnetic moments that are not generated by external
currents in an accretion disk, then they would not possess event horizons.
As noted by Abramowicz, Kluzniak \& Lasota (2002), it is
unlikely that we will ever find direct observational
proof of an event horizon, however, we may be able
to observationally determine whether or not GBHC have intrinsic magnetic
moments. If GBHC are not black holes, they would almost certainly be
magnetized and likely to at least a degree similar to their compact NS cousins.

Although there are
widely studied models for generating magnetic fields in accretion disks, most
can produce equipartition fields at best (Livio \& Pringle 1999), perhaps
at the expense of being too luminous (Bisnovatyi-Kogan \& Lovelace 2000)
in quiescence. While tangled magnetic fields in accretion disks are very likely
responsible for their large viscosity, (e.g. Hawley, Balbus \& Winters 1999)
the highly variable mass accretion rates in LMXB make it unlikely that
disk dynamos could produce the stability of fields needed to account for either
spectral state switches or quiescent spin-down luminosities.
Both also require magnetic fields co-rotating with
the central object. Further, if disk
dynamos produced the much larger apparent magnetic moments of GBHC, they should 
produce them also for the NS systems and cause
profound qualitative spectral and timing differences via interactions 
with the intrinsic NS magnetic moments. Such qualitative
differences as have been observed,
e.g., the hard spectral tail of the high/soft state, lack of surface bursts
for GBHC and stronger GBHC jets, are easily explained by differences 
in masses and magnetic field strengths. Not only are there
are no observed differences that require explanation in terms of event 
horizons (Abramowicz, Kluzniak \& Lasota 2002), there appear to be none
that would be consistent with having two different magnetic structures for
NS and GBHC.

It has been suggested that stable
magnetic fields could be produced by electrically charged, rotating black holes
(Punsley 1998, Gnedin et al. 2003), however the charge necessary to endow Cygnus X-1
with a $10^8$ G magnetic field, well out in the accretion disk,
was found to be $5 \times 10^{28}$ esu (Gnedin et al. 2003). Due to the
large charge/mass ratios of accreting protons or electrons, 
this quantity of black hole charge
would produce electric forces at least $\sim 10^6$ larger than the 
gravitational attraction of $10 M_\odot$,
thus causing charges of one sign to be swallowed and the other to
be blown away. At accretion rates needed to account for the x-ray
luminosity of Cygnus X-1, the original
charge would be neutralized in a fraction of one second. Thus it
appears that current black hole models are unable to offer 
unified explanations of such obviously magnetic
phenomena as jets, spectral state
switches and quiescent synchrotron emissions and if they could, it seems
unusually generous for nature to have provided different mechanisms for
NS and GBHC to produce such strikingly similar phenomena.
 
\section{The Strong Principle of Equivalence}
Astrophysicists nowadays generally
accept the inevitability of the curvature singularities of black holes,
however, if the GBHC are intrinsically magnetized, this will be nature's
way of telling us that such singularities are not really 
permitted to exist. If so, we are left with the task of finding a
fundamental reason for their prohibition. In General Relativity
(GR) the strong principle of equivalence (SPOE) requires that Special Relativity
(SR) must hold locally for freely falling time-like observers in all of spacetime.
This SPOE requirement is a tensor relationship that implies that (i) the
spacetime manifold for observers located in field-free regions, distant
from gravitating masses, must approach the flat spacetime of SR and (ii)
the spacetime world lines of massive matter must always be timelike. 
\footnote{Models of gravitational collapse that lead to
the development of event horizons and central curvature singularities
inevitably abandon the SPOE requirement for timelike world line completeness.
The vanishing of the metric coefficient $g_{tt}$ at the Schwarzschild
radius is sufficient to cause free-fall geodesics of test particles in
non-singular Finkelstein or Kerr-Schild coordinates, for constant central
mass, to become null at the
event horizon, though the crossing can at least be accomplished in a finite
coordinate time, as well as proper time, for Kerr-Schild. 
See Appendix A. It has been shown (Leiter \& Robertson 2003) that
null geodesics occur at surfaces of infinite redshift. In Kruskal-Szekeres
coordinates, in which $g_{tt}$ does not vanish,
there is no surface of infinite redshift at the
Schwarzschild radius and timelike test particle geodesics can traverse it
\textit{in either direction}, so long as the initial conditions are
chosen in a manner that permits the `time' coordinate
to change in a positive sense. However, a central singularity still exists
in these coordinates and they cannot apply to a
gravitational collapse process (Weinberg 1972, Rindler 1977) nor can
they describe the distant asymptotically flat spacetime. They do, however, have
a surface of infinite redshift as $r \rightarrow \infty$, at which outgoing
geodesics become null. Thus they appear
to have no applicability to astrophysics.}
Such spacetime manifolds
are known as `bundle complete' (Wheeler \& Ciuofolini 1995).

As a guiding principle, we look for solutions of the Einstein equations 
that satisfy the SPOE requirement that the world lines of
physical matter under the influence of both
gravitational and non-gravitational forces must not be allowed to become null
or spacelike in any region of spacetime. 
Since the energy-momentum tensor serves as both a source of curvature in the Einstein
equations and a generator of the equations of motion of matter, any constraints
on the latter will affect the former. Hence to assure 
`timelike world line completeness', the
right hand side of the GR field equation
\begin{equation}
G^{\mu\nu}=(8\pi G/c^4) T^{\mu\nu}
\end{equation}
must contain non-gravitational elements capable of stopping the
collapse of physical matter before the formation of a `trapped surface',
thus dynamically avoiding the Hawking and Penrose theorem which states
that once a trapped surface is formed, an event horizon and curvature singularities
are unavoidable. 

In this context, we have found that it is possible to virtually 
stop and maintain a collapsing compact physical plasma object outside of its 
Schwarzschild radius with photon pressure generated by synchrotron radiation from
an equipartition magnetic field, though the object must
then radiate at the local Eddington limit. 
There is strong recent evidence for
the presence of such extreme magnetic fields in gravitational collapse.
Equipartition magnetic fields have been implicated as the driver of
GRB 021206 (Coburn \& Boggs 2003) and strong residual fields much in
excess of those expected from mere flux compression during stellar collapse
have been found in magnetars (Ibrahim, Swank \& Parke 2003).
Kluzniak and Ruderman (1998) have described the generation of $\sim 10^{17}$ G
magnetic fields via differential rotation in neutron stars.
It is likely that larger fields might be generated for the more
massive GBHC. An equipartition field would have $B^2/8\pi \sim \rho c^2/3$,
where $B$ is magnetic field strength and $\rho$ the mass density.
Generating $B \sim 10^{18}$ G, for nuclear densities, should be possible
via differential rotation and relativistic field compression
as surface redshift increases from the vicinity of $z \gtrsim 0.1$
during the final adiabatic decay of
core neutrino emissions. Other possibilities for producing extreme
magnetic fields would include ferromagnetic phase transitions during
the collapse (Haensel \& Bonnazzola 1996, J. Noble private communication)
or the formation of quark condensates.
Fields of this order have been modeled for quark condensates
(Tatsumi 2000), who notes that quark liquids can undergo a ferromagnetic
phase transition at densities as low as nuclear saturation.
An equipartition magnetic field seems to be a
necessary part of a mechanism to ensure the SPOE requirement for
timelike worldline completeness. \footnote{Work by Baumgarte \& Shapiro (2003)
and Thorne (1965) have explored differing cases of magnetic field strength.
The former considers a pressureless dust with matter gravitation dominating the
magnetic field and finds a collapse to a black hole state (that
nevertheless violates timelike world line completeness.) As recounted
by Thorne (1994) dipole-like magnetic fields without rest mass do not
collapse to form trapped surfaces. These results should heighten the
interest in detailed numerical calculations for a radiating equipartition case.}

At the temperatures and compactness of stellar collapse, a pair
plasma is produced. Pelletier \& Marcowith (1998)
have shown that the energy density of magnetic perturbations in
equipartition pair plasmas is preferentially converted to photon pressure,
rather than causing particle acceleration.
The radiative power of an equipartition pair plasma is proportional to
$B^4$, (pair density $\propto B^2$ and synchrotron energy production $\propto
B^2$.) Lacking the equipartition pair plasma, magnetic stress, $B^2/8\pi$,
and gravitational stress, $GM\rho/R$, on mass density $\rho$, would
both increase as $R^{-4}$ during gravitational collapse. Magnetic fields below
equipartition levels would be incapable of stopping the collapse. However,
since photon pressure generated by the pairs at equipartition increases more
rapidly than gravitational stresses, it is possible to stabilize the rate of collapse
at an Eddington limit rate. With this extremely efficient photon production
mechanism, the radiation temperature is buffered near the pair production
threshhold. In effect, the rate of collapse is buffered by a phase transition.
For equipartition conditions, the field also exceeds that required to confine
the pair plasma. A stable rate of collapse is maintained
by increased (decreased) photon
pressure ($\propto B^4$) if the field is increased (decreased) by compresssion
or expansion. Since the photon luminosity is not confined to the core it will
not be trapped, as occurs with neutrinos, however, the radiation should be
thermalized as it diffuses through an optically thick environment.
\textit{To reduce the field to the distantly observed
levels implied by our previous GBHC studies would require the existence
of a red shift of $z \sim 10^8$. Thus we are motivated by the SPOE to look for
solutions of the GR field equations that are consistent with Eddington limited,
highly redshifted, gravitational collapse.} The residual, distantly observed
magnetic moment and extremely faint, redshifted radiations
would be the only things that would distinguish such an object
from a black hole (Abramowicz, Kluzniak \& Lasota 2002).

In Section 3 we show that strict enforcement of the SPOE requirement
for timelike world line completeness leads to a
`no trapped surface' condition. In Section 4 we show that the
gravitational collapse of a magnetospheric, eternally collapsing 
object (MECO) that radiates at its local Eddington limit would continue for a
duration exceeding a Hubble time. In Sections 5 - 10, we examine other physical
characteristics of MECO.

\section{A Strict Interpretation of SPOE Requires MECO}
The simplest form of the energy-momentum tensor, which can
satisfy the SPOE requirement 
of time like world line completeness,
is one which describes a collapsing, radiating plasma 
containing equipartiton magnetic fields which
emit outgoing radiation. Between the extremes of pure magnetic
energy (Thorne 1965) and weakly magnetic, radiation dominated polytropic
gases or pressureless dust (Baumgarte \& Shapiro 2003) there should
be cases where the rate of collapse can be stable.
To first order, in an 
Eddington limited radiation dominated context, these can be described by
the energy momentum tensor:
\begin{equation}
T_{\mu}^{\nu} = (\rho + P/c^2)u_\mu u^\nu - P \delta_\mu^\nu + E_\mu^\nu
\end{equation}
where $E_\mu^\nu = Qk_\mu k^\nu$, $k_\mu k^\mu = 0$ describes outgoing
radiation in a geometric optics approximation, $\rho$ is
energy density of matter and $P$ the pressure. 
Energy momentum tensors corresponding to metrics describing ingoing radiation, 
which are used in many black hole model calculations,
(e.g. Baumgarte \& Shapiro 2003) cannot be used here
because they are incompatible with the $Q > 0$ boundary conditions
associated with collapsing, outwardly radiating objects.  

We choose a comoving interior metric given by
\begin{equation}
ds^2=A(r,t)^2c^2dt^2 - B(r,t)^2dr^2- R(r,t)^2(d\theta^2 + sin^2\theta d\phi^2)
\end{equation}
and an exterior Vaidya metric with outgoing radiation
\begin{equation}
ds^2=(1-2GM/c^2R)c^2du^2 + 2c du dR - R^2(d\theta^2 + sin^2\theta d\phi^2)
\end{equation}
where $R$ is the areal radius and $u=t-R/c$ is the retarded observer time.
Following Lindquist, Schwarz \& Misner (1965), we define
\begin{equation}
\Gamma= \frac{dR}{dl}
\end{equation}
\begin{equation}
U= \frac{dR}{d\tau}
\end{equation}
\begin{equation}
M(r,t)= 4\pi \int_{0}^{r}{\rho R^2 \frac{dR}{dr} {dr}}
\end{equation}
\begin{equation}
\Gamma^2 = (\frac{dR}{dl})^2 = 1 - \frac{2GM(r,t)}{c^2R} + \frac{U}{c}
\end{equation}
where $dl$ is a proper length element in a zero angular momentum
comoving frame, $d\tau$ an
increment of proper time, $U$ is the proper time rate of change of
the radius associated with the invariant circumference of the collapsing mass,
and $M(r,t)$ is the mass enclosed within this radius.
The last two of the relations above have been obtained from the $G_0^0$
component of the field equation (Lindquist, Schwarz \& Misner 1965).
At the boundary of the collapsing, radiating surface, s, we find that the
proper time will be positive definite, as required for 
timelike world line completeness if
\begin{equation}
d\tau_s= \frac{du}{1+z_s} =
    du((1 - \frac{2GM(r,t)_s}{c^2R_s} + \frac{U_s}{c})^{1/2} + \frac{U_s}{c}) > 0
\end{equation}
where $z_s$ is the distantly observed redshift of the collapsing surface.
From Equation (9) we see that in order to avoid a violation of the requirement of 
timelike world line completenes for  $Us < 0$, it is necessary
to dynamically enforce the `no trapped surface condition'\footnote{It
might be argued that there might not be a surface that physically
divides matter from radiation inside a
collapsing massive continuum, however, it has been 
shown (Mitra 2000, 2002, Leiter \& Robertson 2003) 
that Equations (5 - 8) and the $G_0^0$ field
equation in a zero angular momentum comoving frame 
produces the `no trapped surface condition' for any 
interior R(r,t), provided that B(r,t) does not become 
singular at a location where A(r,t) vanishes. However 
this requirement will be satisfied as long 
as timelike world line completeness is maintained by 
photon pressure generated by the equipartition
magnetic field everywhere in the comoving frame.
We can consider any interior location and the radiation flux there
without requiring a joined Vaidya metric. But there will
ultimately be an outer radiating boundary and the required
match to the non-singular 
outgoing exterior Vaidya metric guarantees that there 
will be no metric singularity there.}:
\begin{equation}
\frac{2GM_s}{c^2R_s} < 1
\end{equation}
Since there is nothing in the Einstein tensor $G^{\mu\nu}$
that enforces this condition, we must rely on non-gravitational forces in
$T^{\mu \nu}$ to enforce the SPOE condition of time like world line completeness. 
For the MECO model, we use radiation pressure where
\begin{equation}
Q=\frac{-(dM/du)/4 \pi R^2}{(\Gamma_s + U_s/c)^2}
\end{equation}
At the comoving MECO surface the luminosity is
\begin{equation}
L= 4\pi R^2 Q~~ >0.
\end{equation}
and the distantly observed luminosity is
\begin{equation}
L_{\infty} = -c^2\frac{dM_s}{du} = -c^2\frac{dM_s}{d\tau}(1 + z_s)
\end{equation}

\section{Eddington limited MECO}
Among the various equations associated with the collapse
process there are three proper time differential equations applicable to
a compact collapsing and radiating physical surface. When evaluated on the
physical surface (Hernandez Jr.\& Misner, 1966,
Lindquist, Schwartz \& Misner 1965, Misner 1965, Lindquist,
1966) these equations are:
\begin{equation}
\frac{dU_s}{d\tau} = (\frac{\Gamma}{\rho+P/c^2})_s (-\frac{\partial P}{\partial R})_s
- (\frac{G(M + 4\pi R^3 (P + Q ) / c^2)}{R^2})_s
\end{equation}
\begin{equation}
\frac{dM_s}{d\tau}  =  - (4\pi R^2 P c \frac{U}{c})_s - (L (\frac{U}{c} + \Gamma))_s
\end{equation}
\begin{equation}
\frac{d\Gamma_s}{d\tau}  = \frac{G}{c^4}(\frac{L}{R})_s + \frac{U_s}{c^2} (\frac{\Gamma}{\rho+ P /c^2})_s
(-\frac{\partial P}{\partial R})_s
\end{equation}
In Eddington limited steady collapse, the conditions $dU_s/d\tau =0$ and
$U_s \approx 0$ hold after some time, $\tau_{edd}$, that has elaspsed
in reaching the Eddington limited state. Then
\begin{equation}
\frac{dU_s}{d\tau} = \frac{\Gamma_s}{(\rho + P/c^2)_s}(-\frac{\partial P}{\partial R})_s
  - \frac{GM_s}{R_s^2}  = 0
\end{equation}
Where
\begin{equation}
M_s = (M + 4\pi R^3(P + Q )/c^2)_s
\end{equation}
includes the magnetic field energy in the pressure term and radiant energy in Q.

Equation (17) when integrated over a closed surface
can be solved for the net
outward flow of Eddington limited luminosity through the surface.
Taking the escape cone factor of $27(GM_s/c^2R_s)^2/(1+z_s)^2$ into
account, (See Appendix A) the outflowing
(but not all escaping) surface luminosity, L, would be
\begin{equation}
L_{edd}(outflow)_s  =\frac{4\pi G M_s c R^2(1 + z_{edd})^3}
    {27 \kappa R_g^2}
\end{equation}
where $R_g=GM_s/c^2$ and $\kappa \approx 0.4$ cm$^2 / g$ is the
plasma opacity. (For simplicity, we have assumed here that the
luminosity actually escapes from the MECO surface rather than after
conveyance through a MECO atmosphere and photosphere.
The end result is the same for distant observers.)
However the luminosity $L_s$ which appears in equations (15 - 16) is actually
the net luminosity, which escapes through the photon sphere, and is given by
$L_s = L_{edd}(escape)_s = L_{edd}(outflow) - L_{edd}(fallback) =
L_{Edd,s}-L_{Edd,s}(1-27R_g^2/(R(1+z_{edd}))^2$
Thus in equations 15 and 16, the $L_s$ appearing there is given by
\begin{equation}
L_s= L_{edd}(escape)_s = \frac{4\pi GM(\tau)_s c (1+z_{edd})}{\kappa}
\end{equation}
In this context from (15) we have that
\begin{equation}
c^2\frac{dM_s}{d\tau} = -\frac{L_{edd}(escape)_s}{(1+z_s)^2}
       =  - \frac{4\pi G M(\tau)_s c}{\kappa}
\end{equation}
which can be integrated to give
\begin{equation}
M_s(\tau) = M_s(\tau_{edd}) \exp{((-4\pi G / \kappa c)(\tau - \tau_{edd}))}
\end{equation}
This yield a distantly observed MECO lifetime of
$(1+z_s)\kappa c/4\pi G \sim 5\times 10^{16}$ yr for $z_s \sim 10^8$.
Finally, equation (16) becomes
\begin{equation}
\frac{d\Gamma_s}{d\tau}  =\frac{G}{c^4}\frac{L_{edd,s}}{R_s(\tau_{edd})}
\end{equation}
which, in view of (13) has the solution
\begin{equation}
\Gamma_s(\tau) = \frac{1}{1 + z_s(\tau)} =
(1- \frac{2GM_s(\tau_{edd})}{c^2R_s(\tau)_{edd}})^{1/2}  > 0
\end{equation}
which is consistent with (8) and (10).

\section{MECO Characteristics}
If one naively attributes the Eddington limit luminosity to purely thermal
processes, one quickly finds that the required MECO surface temperatures
would be so high that photon energies would be well beyond the pair production
threshhold and the compactness would assure that photon-photon collisions would
produce numerous electron-positron pairs.
Thus the MECO surface region must be dominated by a pair plasma, and
at a temperature buffered near the $\sim 6\times 10^9 K$ threshhold
for photon-photon collisions to produce pairs (Pelletier \& Marcowith 1998).
As we shall see, an electron-positron pair atmosphere of a MECO is an extremely 
significant structure that conveys radiation from the MECO surface to a zone with a 
much lower red shift and larger escape cone from which it escapes.  In order to 
describe this process computationally within a numerical grid, a radial grid interval
no larger than $\sim 10^{-8}R_g$ would be needed, where $R_g = GM/c^2$ 
is the gravitational radius . Although there have been many numerical
studies of the behavior of collapsing compact objects in GR, to our knowledge
none have indicated that they have sufficient numerical resolution to examine the 
extreme red shift regime associated with MECO nor have they considered the emergent
properties of equipartition magnetic fields and pair plasmas at high red shift.

The strength of the poloidal component of the intrinsic magnetic fields $B_{env}$
observed in the distant environment around the MECO are reduced by a factor of $(1+z_s)$
from their values near $R_s$. The fields needed to produce jets
in AGN are observed to be of the order $10^3 - 10^4$ gauss as judged distantly.
On the other hand, a distantly observed equipartition field would
be $\sim 10^{18}/(m(1+z_{edd}))$ gauss,
where $m=M/M_\odot$. This suggests that for an
$m \sim 10^8$ AGN, the combined effect of mass scaling and red shift would need
to reduce the surface field from $10^{18}$ gauss to $10^{3 - 4}$ gauss. This
would require the MECO to have a red shift of $z \sim 10^7 - 10^8$.
In this and previous work, (Robertson \& Leiter 2002) we have found typical
magnetic fields of a few times $10^{10}$ gauss for GBHC. These would
require similar values of $z_{edd} \sim 10^8$, as well as $ m\sim 10$.
Therefore for both GBHC and AGN we find that we need
\footnote{An additional point of support for very large values of z
concerns neutrino transport in stellar core collapse. If a diffusion limited
neutrino luminosity of $\sim 10^{52}$
erg/s (Shapiro \& Teukolsky 1983) were capable
of very briefly sustaining a neutrino Eddington limit
rate of collapse, then the subsequent reduction of
neutrino luminosity as neutrino emissions are depleted in the core would lead to
an adiabatic collapse, magnetic flux compression, and photon emissions 
reaching an Eddington limit. At this point the photon luminosity
would need to support a smaller diameter and more tightly gravitationally
bound mass. A new photon Eddington
balance would thus require an escaping luminosity reduced by at least
the $\sim 10^{20}$ opacity ratio $(\sigma_T/\sigma_\nu)$,
where $\sigma_T = 6.6\times 10^{-25}$ cm$^2$ is the Thompson cross section and
$\sigma_\nu = 4.4\times 10^{-45}$ cm$^2$ is the neutrino scattering
cross-section. Thus $L_\infty \lesssim 10^{31-32}$ erg/s would be
required. For this to be $L_{edd,\infty}$ for a
10 M$_\odot$ GBHC would require $1+z \sim 10^8$. The adiabatic
relaxation of neutrino support and formation of a pair plasma
is an important step in gravitational
collapse that is not encompassed by polytropic equation of state
models of collapse. It is of some interest that if neutrinos have
non-zero rest mass they might
be trapped inside the photon sphere anyway.}
\begin{equation}
1 + z_{edd} = \frac{B_{equip}}{B_{env}} \sim  10^8.
\end{equation}

\section{The quiescent MECO}
The quiescent luminosity of a MECO originates deep within its photon sphere.
When distantly observed it is diminished by both gravitational red shift
and a narrow exit cone. The gravitational red shift
reduces the surface luminosity by $1/(1+z)^2$ while the exit cone further
reduces the luminosity by the factor
$27 R_g^2/(R(1+z))^2 \sim 27/(4(1+z)^2)$ for large z. (See Appendix A).
Here we have used
\begin{equation}
\frac{R_g}{R} = \frac{1}{2}(1 - \frac{1}{(1+z)^2})
\end{equation}
where R and z refer to the location from which photons escape.
The net outflow fraction of the luminosity  provides the support for
the collapsing matter, thereby dynamically maintaining the SPOE requirement 
of timelike world line completeness. The photons which finally escape do so
from the photosphere of the pair atmosphere. The fraction of
luminosity from the MECO surface
that escapes to infinity in Eddington balance is
\begin{equation}
(L_{edd})_s = \frac {4\pi G M_s c(1+z)}{\kappa} = 1.27 \times 10^{38}m(1+z_s)~~~~ erg/s
\end{equation}
The distantly observed luminosity is:
\begin{equation}
L_\infty = \frac{(L_{edd})_s}{(1+z_s)^2} = \frac {4\pi G M_s c}{\kappa(1+z_s)}
\end{equation}
When radiation reaches the photosphere, where the temperature is $T_p$,
the fraction that escapes to be distantly observed is:
\begin{equation}
L_\infty = \frac{4 \pi R_g^2 \sigma T_p^4 27}{(1+z_p)^4}
    = 1.56\times 10^7 m^2 T_p^4 \frac{27}{(1+z_p)^4}~~~~erg/s
\end{equation}
where $\sigma = 5.67\times 10^{-5}$ erg/s/cm$^2$
and subscript p refers to conditions at the photosphere.
Equations 28 and 29 yield:
\begin{equation}
T_\infty = T_p/(1+z_p) = \frac{2.3\times 10^7}{(m(1+z_s))^{1/4}}~~~~ K.
\end{equation}
To examine typical cases, a $10 M_\odot$, $m = 10$ GBHC modeled in
terms of a MECO with $z \sim 10^8$ would have
$T_\infty = 1.3\times 10^5 K = 0.01$ keV, a bolometric luminosity,
excluding spin-down contributions, of
$L_\infty =1.3\times 10^{31} erg/s$, and a spectral peak at 220 A$^0$,
in the photoelectrically absorbed deep UV.
For an m=$10^7$ AGN, $T_\infty = 4160 K$, $L_\infty = 1.3\times 10^{37} erg/s$
and a spectral peak in the near infrared at 7000 A$^o$.
(Sgr A$^*$ at $m \approx 3\times 10^6$, would have $T_\infty =5500$ K,
and a 2.2 micron brightness of 6 mJy, just below the observational
upper limit of 9 mJy (Reid et al. 2003).)
Hence passive MECO without active accretion disks, although
not black holes, have lifetimes much greater
than a Hubble time and emit highly red
shifted quiescent thermal spectra that may be quite
difficult to observe. There are additional power law components
of similar magnitude that originate as magnetic dipole
spin-down radiation (see below).

Escaping radiation passes through a pair plasma atmosphere that can be shown,
\textit{ex post facto} (See Appendix B), to be radiation dominated throughout.
Under these circumstances, the radiation pressure within the equilibrium
atmosphere obeys $P_{rad}/(1+z) = constant$. \footnote{Due to its
negligible mass, we
consider the pair atmosphere to exist external to the Meco. Due to the slow
collapse, the exterior Vaidya metric can be approximated by exterior,
outgoing Finkelstein coordinates. In this case, the hydrostatic balance
equation within the MECO atmosphere is
$\frac{\partial p}{\partial r} =
-\frac{\partial \ln{(g_{00})}}{2 \partial r}(p + \rho c^2)$, where
$g_{00} = (1-2R_g/r)$ and $\rho c^2 << p$. This integrates to $p/(1+z) = constant$.}
Thus the relation between surface and photosphere temperatures is
$T_s^4/(1+z_s) = T_p^4/(1+z_p)$. At the MECO surface, we expect a
pair plasma temperature of $T_s \approx m_ec^2/k \sim 6\times 10^9 K$ because
an equipartition magnetic field effectively acts as a thermostat which
buffers the temperature of the optically thick synchrotron radiation
escaping from the MECO surface (Pelletier \& Marcowith 1998).
But since $T_\infty = T_p/(1+z_p)$,
we have that
\begin{equation}
T_p = T_s(\frac{T_s}{T_\infty (1+z_s)})^{1/3}
= 1.76\times 10^9(\frac{m}{1+z_s})^{1/12} ~~~~K
\end{equation}
For a $m=10$ and $1+z_s=10^8$ GBHC, this yields a photosphere temperature of
$4.6\times 10^8$ K, from which $(1+z_p) = 3500$. An AGN with $m=10^7$
would have a somewhat warmer photosphere at $T_p = 1.5\times 10^9$ K, but
with a red shift of 360000. 

\section{An Actively Accreting MECO}
From the viewpoint of a distant observer, accretion would
deliver mass-energy to the MECO, which would then radiate most of it
away. The contribution from the central MECO alone would be
\begin{equation}
L_\infty = \frac {4\pi G M_s c}{\kappa(1+z_s)}+ \frac{\dot{m}_\infty c^2}{1+z_s}(e(1+z_s)-1)
    = 4 \pi R_g^2 \sigma T_p^4 \frac {27} {(1+z_p)^4}
\end{equation}
where $e = E/mc^2 = 0.943$ is the specific energy per particle
available after accretion disk flow to the marginally stable orbit radius,
$r_{ms}$. Assuming that $\dot{m}_\infty$ is some fraction, f, of the
Newtonian Eddington limit mass accretion rate, $4\pi G M c/\kappa$, then
\begin{equation}
1.27\times 10^{38}\frac{m\eta}{1+z_s} =
(27)(1.56\times 10^7)m^2(\frac{T_p}{1+z_p})^4
\end{equation}
where $\eta=1+f((1+z_s)e -1)$ includes both quiescent and accretion
contributions to the luminosity.
Due to the extremely strong dependence on temperature of the
density of pairs, (see Appendix B) it is unlikely that the temperature
of the photosphere will be greatly different from the average of $4.6\times 10^8
K$ found previously for a typical GBHC. Assuming this to be the case, along with
$z=10^8$, $m=10$, and $f=1$, we find $T_\infty = T_p/(1+z_p) = 1.3\times 10^7 K$
and $(1+z) = 35$, which indicates considerable photospheric expansion.
The MECO luminosity would be approximately Newtonian Eddington limit at
$L_\infty = 1.2\times 10^{39}$ erg/s. For comparison, the accretion disk
outside the marginally stable orbit at $r_{ms}$
(efficiency = 0.057) would produce only $6.8\times 10^{37}$ erg/s.
Thus the high accretion state luminosity of a GBHC would originate
primarily from the central MECO. The thermal component
would be `ultrasoft' with a temperature of only $1.3\times 10^7 K$ (1.1 keV).
A substantial fraction of the
softer thermal luminosity would be Compton scattered to higher energy
in the plunging flow inside $r_{ms}$. Even if a disk flow
could be maintained all the way to the MECO
surface, where a hot equatorial band might result, the escaping
radiation would be spread over the larger area of the photosphere due
to photons origins deep inside the photon orbit.

For radiation passing through the photosphere
most photons would depart with some azimuthal momentum on spiral trajectories
that would eventually take them across and through the accretion disk.
Thus a very large fraction of the soft photons would be subject
to bulk comptonization in the plunging region inside $r_{ms}$.
This contrasts sharply with the situation for neutron stars
where there probably is no comparable plunging region and
few photons from the surface cross the disk. This
could account for the fact that hard x-ray spectral tails are
comparatively much stronger for high state GBHC.
Our preliminary calculations for photon trajectories
randomly directed upon leaving the
photon sphere indicate that this process
would produce a power law component with photon index greater than 2.
These are difficult, but important calculations for which the effects
of multiple scattering must be considered. But they are beyond the
scope of this paper, which is intended as a first description of the
general MECO model.

\section{Spectral States}
The progression of configurations of accretion disk, magnetic field
and boundary layer is shown in Figure 1. The caption summarizes
the spectral features expected in four regimes:

{\bf Quiescence}\\
A low accretion rate is ablated by $\sim 10^{30-33}$ erg/s
radiation from the central object at a large inner disk radius.
This luminosity is sufficient to raise the temperature of the optically
thick inner disk above the $\sim 5000$ K instability temperature for
hydrogen out to a distance of $r \sim 10^{10}$ cm. Therefore we expect the
quiescent inner disk to be essentially empty with a large inner radius.
The rate of mass flow ablated at the inner disk radius would only
need to be $\sim 10^{13}$ g/s to produce the
quiescent optical emission observed for GBHC and NS. The ablated material
could escape if it reached the magnetic propeller region, which is
confined to the light cylinder at a much smaller radius, $r_{lc}$,
than that of the inner disk. This makes the MECO model
compatible with the disk instability model of x-ray nova
outbursts, which begin as `outside-in' events in which substantial outer mass
reservoirs have been observed to  fill an accretion disk on the
viscous timescale of a very subsonic radial flow (Orosz et al. 1997).

Quiescent luminosities that are generally 10 - 100 X lower for GBHC
than for neutron stars (NS) have been claimed as evidence for the existence
of event horizons. (Narayan et al. 1997, Garcia et al. 2001).
In the MECO model, the quiescent emissions are magnetic dipole
emissions that are characteristic of the magnetic moment and rate of
spin of the central object. The lower quiescent luminosities of the
GBHC are explained by their lower spin rates and (perhaps unobservably) low
rates of quiescent emission from the central MECO.

{\bf Low State}\\
The inner disk radius is inside the light cylinder, with hot, diamagnetic
plasma reshaping the magnetopause topology (Arons et al. 1984). This magnetic
propeller regime (Ilarianov \& Sunyaev 1975, Stella, White
\& Rosner 1986, Cui 1997, Zhang, Yu \&
Zhang 1997, Campana et al. 1998) exists until
the inner disk pushes inside the co-rotation radius, $r_c$.
At that time, large fractions of the accreting plasma can continue
on to the central object and produce a spectral state switch to softer
emissions. In previous work (Robertson \& Leiter 2002) we found that
magnetic moments and spin rates could be determined from luminosities
at the end points of the spectral hardening transition.
The magnetic moments and spins
were used to calculate the $\sim 10^{3-6}$ times fainter quiescent luminosities
expected from spin-down. The results are recapitulated and extended
in Table 1 and Appendix D.
Calculated values of quiescent
luminosity in Table 1 have been corrected using
a more recent correlation of spin-down energy loss rate and soft
x-ray luminosity (Possenti et al. 2002), but results are otherwise unchanged
from the previous work except for new additions listed in bold font.
It is very powerful confirmation of the propeller mechanism that spins are
in good agreement with burst oscillation frequencies (Strohmayer \&
Markwart 2002, Chakrabarty et al. 2003),
magnetic moments are of similar magnitude to
those determined from the spin-down of similarly rotating
millisecond pulsars and the calculated quiescent luminosities are accurate.

\begin{table*}
\begin{center}
\caption{$^a$Calculated and Observed Quiescent Luminosities}
\begin{tabular}{lrrrrrrrr} \hline
Object & m & $L_{min}$ & $L_c$ & $\mu_{27}$ & $\nu_{obs}$ & $\nu_{calc}$ & log (L$_q$) & log (L$_q$) \\
    & M$_\odot$ & $10^{36}$erg/s & $10^{36}$erg/s & Gauss cm$^3$    & Hz      &  Hz~ & erg/s & erg/s \\
    & & & & & obs. & calc. & obs. & calc. \\ \hline
\bf {NS} \\
Aql X-1 & 1.4 & 1.2 & 0.4 & 0.47 & 549 & 658 & 32.6  & 32.5  \\
4U 1608-52& 1.4 & 10 &2.9 & 1.0 & 619 & 534 & 33.3 & 33.4 \\
Sax J1808.4-3658& 1.4 & $^b$0.8 & 0.2 & 0.53 & 401 & 426 & 31.8-32.2 & 32 \\
Cen X-4 & 1.4 & 4.4 & 1.1 & 1.1 & & 430 & 32.4 & 32.8 \\
KS 1731-26 & 1.4 & & 1.8 & 1.0 & 524 & & $^c$\bf{32.8} & 33.1  \\ 
\bf{XTE J1751-305} & 1.4 & & 3.5 & 1.9 & 435 & & $<$34.3 & 33.7 \\
\bf{XTE J0929-314} & 1.4 & & 4.9 & 8.5 & 185 & & & 33.1 \\
4U 1916-053 & 1.4 & $\sim$14 & 3.2 & 3.7 & 270 & 370 & & 33.0 \\
\bf{4U1705-44} & 1.4 & 26 & 7 & 2.5 & & 470 & & 33.7 \\
4U 1730-335 & 1.4 & 10 & & 2.5 & 307 & & & 32.9 \\
\bf{GRO J1744-28} & 1.4 & & 18 & 13000 & 2.14 & & & 31.5 \\
Cir X-1 & 1.4 & 300 & 14 & 170 & & 35 &  & 32.8 \\ \hline
\bf{GBHC} \\
GRS 1124-68 & 5 & 240 & 6.6 & 720 & & 16 & $<32.4$ & 32.7 \\
GS 2023+338 &7 & 1000 & 48 & 470 & & 46 & 33.7 & 34 \\
XTE J1550-564 & 7 & $^d$90 & 4.1 & 150 & & 45 & 32.8 & 32.2 \\
GS 2000+25 & 7 & & 0.15 & 160 & & 14 & 30.4 & 30.5\\
GRO J1655-40 & 7 & 31 & 1.0 & 250 & & 19 & 31.3 & 31.7 \\
A0620-00 & 4.9 & 4.5 & 0.14 & 50 &  & 26 & 30.5 & 30.2 \\
Cygnus X-1 & 10 & & 30 & 1260 & & 23 &  & 33 \\
GRS 1915+105 & 7 & & 12 & 130 & $^e$67 & & & 33 \\
\bf{XTE J1118+480} & 7 & & 1.2 & 1000 & & 8 & & 31.5 \\
\bf{LMC X-3} & 7 & 600 & 7 & 860 & & 16 & & 33 \\ \hline
\end{tabular}
\end{center}
$a$New table entries in bold font are described in Appendix D. \\
Equations used for calculations of spins, magnetic moments and $L_q$ are in Appendix D.\\
Other tabular entries and supporting data are in Robertson \& Leiter (2002)\\
$^b$2.5 kpc, $^c$ (Burderi et al. 2002), $^d$d = 4 kpc \\
$^e$GRS 1915+105 Q $\approx 20$ QPO was stable
for six months and a factor of five luminosity change. \hfill \\
\end{table*}
Plasma flowing outward in the low state may depart in a jet, or as an outflow
back over the disk as plasma is accelerated on
outwardly curved or open magnetic field
lines. Radio images of both flows have been seen (Paragi et al. 2002).
Equatorial outflows could contribute to the low state hard spectrum
by bulk Comptonization of soft photons in the outflow. This would
accentuate the hardness by the depletion of the soft photons that would
otherwise be observed to arise from the disk. Such an outflow
would be compatible with partial covering models for dipping sources, in which
the hard spectral region seems to be extended
(Church 2001, Church \& Balucinska-Church 2001).
Alternatively, an accretion disk corona or compact jet
might be a major contributor to
the hard spectrum. For jet emissions, recent work (Corbel
\& Fender 2002) has shown that it may be possible to explain much of the
broadband emissions from near infrared through soft x-rays as
the power-law synchrotron emissions of compact jets, which
have been directly imaged for some GBHC.

{\bf Intermediate State}\\
Intermediate states occur when some, but not all, of the accreting matter
can make its way to the central object. Incomplete spectral state
switches terminating well below the Eddington limit,
such as those exhibited by Cygnus X-1 may occur. If the co-rotation
radius is large, there is a very large difference in the efficiency of
energy release at the central object vs the disk. Thus changes
of luminosity and an apparent spectral state switch can occur for very
small change of accretion rate.
Relaxation oscillations driven by
intermittent radiation from the central object can occur if the
accretion rate is not steady. Large periodic jet ejections may be
associated with this state, for which significant toroidal winding
of the poloidal magnetic field lines and radiation pressure may contribute
to the ejection (see below).

{\bf High State}\\
With the disk inner radius inside $r_c$, the
propeller regime ends and matter of sufficient pressure can make its way inward.
From quiescence to the light cylinder, the
x-ray luminosity changes by a factor of only a few as the disk
generates a soft thermal spectral component (which may be mistaken
for surface radiation for NS.) From $r_{lc}$ to $r_c$, the x-ray luminosity
may increase by a factor of $\sim 10^3 - 10^6$. With inner disk inside
$r_c$, the outflow and/or jets subside, the system becomes
radio quiet, the photon index increases, and a soft
thermal excess from the central object appears, both of which
contibute to a softer spectrum, (e.g., see Fig. 3.3 of Tanaka \& Lewin, p. 140),
which may be even be described as `ultrasoft' (White \& Marshall 1984);
particularly when the central object cools as
the luminosity finally begins to decline. We have
shown that MECO would produce a dominant `ultrasoft' component in the
high state. They would also produce an extensive power-law hard tail
as soft photons leaving the MECO well inside the photon orbit take
trajectories that take them across the plunging region inside the
marginally stable orbit. Since there is no comparably rich source of photons
on disk crossing trajectories for NS and a much smaller, if any, plunging
region, there is no comparable hard spectral component in their high states.
Another mechanism that might contribute to the power-law tail is
cyclotron resonance scattering in the outer photosphere (Thompson, Lyuitikov
\& Kulkarni 2002). In either case, the high state hard tail has different
origins and characteristics from the hard tail of the low state.

\section{Disk Characteristics}
For matter suffiently
inside $r_c$, the propeller mechanism is incapable of stopping the flow,
however, a boundary layer may form at the inner disk radius in this case.
The need for a boundary layer for GBHC can be seen by comparing the magnetic
pressure at the magnetosphere with the impact pressure of a trailing,
subsonic disk. For example, for an average GBHC magnetic moment of
$\sim 4\times10^{29}$ gauss cm$^3$ from Table 1, the magnetic pressure
at a $r_{ms}$ radius of $6.3\times10^6$ cm for a 7 M$_\odot$
GBHC would be $B^2 /8\pi \sim 10^{17}$ erg/cm$^3$. At a mass flow rate of
$\dot{m} = 10^{18}$ g/s, which would be near Eddington limit
conditions for a 7 M$_\odot$ MECO, the inner disk temperature would be
$T \sim 1.5\times10^7$ K. The disk scale height would be given by
$H\sim r v_s/v_K \sim 0.0036r$, where $v_s \sim 4.5\times 10^7$cm/s
and $v_K \sim 1.2\times 10^{10}$ cm/s are acoustic and
Keplerian speeds, respectively. The impact pressure
would be $\dot{m}v_r/4\pi r H \sim 5.6\times10^5 v_r$ erg/cm$^3$.
It would require $v_r$ in excess of the speed of light to
let the impact pressure match the magnetic pressure.
But since the magnetic field doesn't eject the
disk material inside $r_c$, matter piles up as essentially dead weight against
the magnetopause and pushes it in. The radial extent of such a layer
would only need to be $\sim kT/m_pg \sim 50$ cm,
where $m_p$ is the proton mass and
$g$, the radial gravitational free fall acceleration, but it is likely
distributed over a larger transition zone from co-rotation with the
magnetosphere to Keplerian flow. The gas pressure at the inner radius of the
transition zone necessarily matches the magnetic
pressure. In this case, radiation pressure in the disk,
at $T= 1.5\times 10^7 K$,
is nearly three orders of magnitude below the gas pressure. Therefore a
gas pressure dominated, thin, Keplerian disk with subsonic radial speed
should continue all the way to
$r_{ms}$ for a MECO. Similar conditions occur with disk radius inside $r_c$
even for weakly magnetic `atoll' class NS. The similar magnetic pressures
at $r_c$ for GBHC and atolls is one of the reasons for their spectral and
timing similarities.
The nature of mass accumulations in the inner disk
transition region and the way that they can enter the magnetosphere
have been the subject of many studies, (e.g., Spruit \& Taam 1990).

In the case of NS, sufficiently high mass accretion rates
can push the magnetopause into the star surface, but this
requires near Eddington limit conditions. At this point the
hard apex of the right side of the horizontal branch of the `Z'
track in the hardness/luminosity diagram is reached. It has recently been
shown (Muno et al. 2002) that the distinction between `atoll' and
`Z' sources is merely that this point is reached near the Eddington
limit for `Zs' and at perhaps $\sim 10 - 20$\% of this luminosity
(Barrett \& Olive 2002) for the less strongly magnetized `atolls'.
Atolls rarely reach such luminosities. For MECO based GBHC, one
would expect a relatively constant ratio of hard and soft x-ray
`colors' after the inner disk crosses $r_c$ and the
flow reaches the photon orbit. If x-ray `color' bands for GBHC were chosen
below and above a $\sim 1 keV$ thermal peak similarly to way they
are now chosen to bracket the $\sim 2 keV$ peak of NS, one might observe
a 'Z' track for the color/color diagrams of GBHC.

An observer at coordinate, r, inside $r_{ms}$, would find the
radial infall speed to be $v_r = \frac{\sqrt{2}}{4}c(6R_g/r-1)^{3/2}$,
(see Appendix A) and the Lorentz factor for a particle spiraling
in from $6R_g$ would be $\gamma = 4 \sqrt{2}(1 + z)/3$, where
$1+ z = (1-2R_g/r)^{-1/2}$ would be the red shift for photons generated at $r$.
If the distantly observed mass accretion rate would be $\dot{m}_\infty$, then
the impact pressure at r would be $p_i = (1+z)\dot{m}_\infty \gamma v_r/ (4\pi r H)$.
For $\dot{m}_\infty \sim 10^{18}$ g/s, corresponding to Eddington limit
conditions for a 7 M$_\odot$ GBHC, and $H = 0.0036r$, impact pressure is,
$p_i \sim 5 \times 10^{16}(1 + z)^2(2R_g/r)^2(6R_g/r-1)^{3/2}$
erg/cm$^3$. For comparison, the magnetic pressure
is $(1 + z)^2 B_\infty^2/8\pi$. Assuming a dipole
field with average magnetic moment of $4\times 10^{29}$ gauss cm$^3$
from Table 1, the magnetic
pressure is $\sim 10^{20} (1+z)^2 (2R_g/r)^6$ erg/cm$^3$.
Thus there are no circumstances for
which the impact pressure is as large as the
magnetic pressure for $2R_g < r < 6R_g$.
We conclude that another weighty boundary layer must form inside
$r_{ms}$ if the magnetosphere is to be pushed inward.
More likely, the plasma stream is broken up by Kelvin-Helmholtz
instabilities and filters through the magnetosphere. In any event,
the inner radius of the disk is determined by the rate at which
the magnetic field can strip matter and angular momentum from the disk.
This occurs in a boundary layer of some thickness, $\delta r$, that
is only a few times the disk thickness. (See Appendix C)

Other than the presence of a transition boundary layer on the magnetopause,
the nature of the flow and spectral formation inside $r_c$ is a research topic.
Both the short radial distance from $r_c$ to $r_{ms}$ and the
magnetopause topology should help to maintain a
disk-like flow to $r_{ms}$. Radial acceleration inside $r_{ms}$
should also help to maintain a thin flow structure. These flows are depicted
in Figure 1. Recapitulating, we expect the flow into the
MECO to produce a distantly observed soft thermal component, part of which is
strongly bulk Comptonized.

{\bf Quasi-periodic Oscillations}\\
Although many mechanisms have been proposed for the high frequency
quasi-periodic oscillations (QPO) of x-ray luminosity, they
often require conditions that are incompatible with thin, viscous
Keplerian disks. Several models have requirements for
lumpy flows, elliptical inner disk boundaries,
orbits out of the disk plane or conditions
that should produce little radiated power. In a conventional thin disk,
the vertical oscillation frequency, which is approximately the same
as the Keplerian frequency of the inner viscous disk radius should
generate ample power.  Accreting plasma should periodically wind the poloidal
MECO magnetic field into toroidal configurations until the field lines break
and reconnect across the disk. Field reconnection across the
disk should produce high frequency oscillations that couple to the
vertical oscillations. If so, there would be an automatic association of high
frequency QPO with the harder power-law spectra of magnetospherically driven
emissions, as is observed. Mass ejection in low state jets might be related to
the heating of plasma via the field breakage mechanism, in addition
to natural buoyancy of a plasma magnetic torus in a poloidal external field.

It seems possible that toroidal winding and reconnection
of field lines at the magnetopause, might
continue in high states inside $r_{ms}$.
If so, there might be QPO that could be identified as signatures of
the MECO magnetosphere. If they occur deep within the magnetosphere, they
might be at locally very high frequencies, and be observed distantly
as very redshifted low frequencies. As shown in Appendix A, the
`Keplerian' frequencies in the plunging region inside $r_{ms}$ are
given by $\nu = 1.18\times 10^5 (R_g/r)^2(1-2R_g/r)/m$ Hz. A maximum frequency of
437 Hz would occur for m=10 at the photon orbit. Of more interest, however
are frequencies for $R_g/r \approx 1/2$, for which $\nu = 2950/(m(1+z)^2)$ Hz.
For $1+z = 10 - 100, m=10$; conditions that might apply to the photosphere
region, $\nu \sim 0.03 - 3$ Hz could be produced. In this regard, one could
expect significant time lags between inner disk accretion and
luminosity fluctuations and their echoes from the central highly
redshifted MECO.

Even if QPO are not produced inside $r_{ms}$ or inside the photon sphere
for GBHC, there is an interesting scaling mismatch that might allow them
to occur for AGN. Although the magnetic moments of AGN scale inversely
with mass, the velocity of plasma inside $r_{ms}$ does not. Thus the
energy density of disk plasma inside $r_{ms}$ will be relatively larger than
magnetic field energy densities for AGN accretion disks. When field
energy density is larger than kinetic energy density of matter, the field
pushes matter around. When the reverse is true, the matter drags the field
along. Thus toroidal winding of the field at the magnetopause
could fail to occur for GBHC, but might easily do so for AGN.
If the process is related to mass ejection, then very energetic jets
with Lorentz factors $\gamma \sim (1+z)$ might arise from within
$r_{ms}$ for AGN. A field line breakage model of `smoke ring'
like mass ejection from deep within $r_{ms}$ has been developed by
Chou \& Tajima (1999). In their calculations, a pressure of unspecified
origin was needed to stop the flow outside $2R_g$ and a poloidal
magnetic field, also of unspecified origin was required. MECO
provide the necessary ingredients in the form of the intrinsic
MECO magnetic field. The Chou \& Tajima mechanism,
aided by intense radiation pressure, may be active
inside $r_{ms}$ for GBHC and produce extremely large episodic mass
ejections such as those shown by GRS 1915+105. Although not
developed for conditions with large inner disk radius, the same
magnetic mechanism could produce the jet emissions
associated with the low/hard state (Gallo, Fender \& Pooley 2003).

Finally, some of the rich oscillatory behavior of GRS 1915+105 may be
readily explained by the interaction of the inner disk and the central
MECO. The objects in Table 1 have co-rotation radii of order $20 R_g$,
which brings the low state inner disk radius
in close to the central object. A low state MECO, balanced near
co-rotation would need only a small increase of mass flow rate to
permit mass to flow on to the central MECO. This would produce more than
20X additional luminosity and enough radiation pressure to blow the
inner disk back beyond $r_c$ and load its mass onto the magnetic field
lines where it is ejected. This also explains the association of jet
outflows with the oscillatory states. Belloni et al. (1997) have shown
that after ejection of the inner disk, it
then refills on a viscous time scale until the process repeats.
Thus one of the most enigmatic GBHC might be understood as a relaxation
oscillator, for which the frequency is set by a critical
mass accretion rate.

\section{Detecting MECO}
It may be possible to detect MECO in several ways. Firstly, as we
have shown, for a red shift
of $z \sim 10^8$, the quiescent luminosity of a GBHC MECO would be
$\sim 10^{31}$ erg/s with $T_\infty \sim 0.01$ keV. This thermal peak might be
observable for nearby  or high galactic latitude GBHC,
such as A0620-00 or XTE J1118+480. 
Secondly, at moderate luminosities $L \sim 10^{36} - 10^{37}$ erg/s but
in a high state at least slightly above $L_c$,
a central MECO would be a bright, small central object that might
be sharply eclipsed in deep dipping sources. A high state MECO should stand out
as a small bright source. This is consistent with analyses of absorption dips in 
GRO J1655-40 (Church 2001) which have shown the soft source of the high state to
be smaller than the region that produces the hard spectral component of its low states.
A conclusive demonstration that most of the soft x-ray
luminosity of a high state GBHC is distributed over a large accretion disk would be
inconsistent with MECO or any other GBHC model entailing a central
bright source. If the MECO model is correct, the usual identification of the
bright, high state soft component as disk emissions would be wrong. 
Thirdly, a pair plasma atmosphere in an
equipartition magnetic field should be virtually transparent to photon
polarizations perpendicular to the magnetic field lines. The x-rays
from the central MECO should exhibit some polarization that might be
detectable, though this is far from certain since the distantly observed
emissions could originate from nearly any point on the photosphere.
MECO presumably would not be found only in binary systems.
If they are the offspring of massive
star supernovae, then they should be found all over the galaxy. If we
have correctly estimated their quiescent temperatures, isolated MECO
would be weak, possibly polarized, EUV sources with a
power-law tail in soft x-rays.

\section{Conclusions}
It is now becoming apparent that many of the spectral properties of
LMXB, including the GBHC, are consistent with
the existence of intrinsically magnetized central objects. We have shown that the 
existence of intrisically magnetic GBHC are consistent with a new class of 
magnetospheric eternally collapsing object (MECO) solutions of the Einstein field 
equations of General Relativity. These solutions are based on a strict
adherence to the Strong Principle of Equivalence (SPOE) requirement for timelike 
world line completeness; i.e., that the world lines of physical 
matter under the influence of gravitational and non-gravitational forces must 
remain timelike in all regions of spacetime.
Since there is nothing in the structure of the Einstein tensor, $G^{\mu \nu}$,
on the left hand side of the Einstein field equation that dynamically enforces
`time like world line completeness', we have argued that the SPOE constrains the 
physically acceptable choices of the energy momentum tensor, $T^{\mu \nu}$
to contain non-gravitational forces that can dynamically enforce it. In this context 
we have found the MECO solutions. Since MECO lifetimes are orders of magnitude 
greater than a Hubble time, they provide an elegant and unified framework for 
understanding the broad range of observations associated with GBHC and active 
galactic nuclei. 

An enormous body of physics scholarship developed primarily over the last half 
century has been built on the assumption that trapped surfaces leading to event 
horizons and curvature singularities exist. Misner, Thorne \& Wheeler (1973), for 
example in Sec. 34.6 clearly state that this is an assumption and that it underlies 
the well-known singularity theorems of Hawking and Penrose. In contrast, we have 
found that strict adherence to the SPOE demand for timelike world line completeness 
requires a \textit{`no trapped surface condition'}. This has led to the 
quasi-stable, high red shift MECO solutions of the Einstein field equations.
The physical mechanism of their stable rate collapse is an Eddington balance maintained 
by the distributed photon generation of an equipartition magnetic field. 
This field also serves to confine the pair plasma dominated outer layers of 
the MECO and the thin MECO pair atmosphere. Red shifts of $z \sim 10^8$ have 
been found to be necessary for compatibility with our previously found 
magnetic moments for GBHC. 

We have delineated the expected spectral properties of MECO in
quiescence and in accreting states and shown that the central magnetic moments
are large enough to require their accretion disks to be dominated
by gas pressure, though the inner disk may be subject to substantial
irradiation. The MECO model seems able to
robustly account for the general spectral and timing
behaviours of GBHC while providing for
their appropriate quantitative differences from NS. Lastly, we have
indicated some ways in which MECO might be detected and confirmed.

\acknowledgements


\appendix{\bf A. Relativistic particle mechanics}\\
A number of standard, but useful results for relativistic mechanics are
recapitulated here. All are based upon the energy-momentum four-vector
for a free particle in the singularity-free Finkelstein or Kerr-Schild
coordinates for a constant central mass. Though not strictly
compatible with radiating objects with variable mass, outgoing Finkelstein
coordinates are a useful first order approximation to the 
outgoing Vaidya coordinates for low radiation rates exterior to a MECO.
\begin{equation}
ds^2=dt^2((1-2R_g/r) \pm 4R_g v^r/r - (1+2R_g/r)v^r v^r) -r^2(d\theta^2+\sin \theta^2 d\phi^2)
\end{equation}
The plus sign corresponds to outgoing Finkelstein coordinates and
the negative sign to ingoing Finkelstein or Kerr-Schild coordinates.
Here $v^r=dr/dt$.
For a particle in an equatorial trajectory ($\theta$ = $\pi$,
$p_{\theta} = 0$) about
an object of gravitational mass M, one obtains the same equation as
for Schwarzschild coordinates:
\begin{equation}
(\frac{dr}{d\tau}) = -c(e^2-(1-2R_g/r)(1+a^2(R_g/r)^2))^{1/2}
\end{equation}
Where $e$ is the conserved energy per unit rest mass,
$a=(cp_\phi/GMm_0)$ is a dimensionless, conserved angular momentum,
$\tau$ is the proper time in the particle frame and the negative sign
indicates movement toward $r=0$.
The metric Equation (1) also describes these
radial geodesics with $ds^2 = d\tau^2$
Neglecting angular terms and letting $q=dt/d\tau$,
this equation can be written as
\begin{equation}
1=(1-2R_g/r)q^2 \pm 4pqR_g/r - (1+2R_g/r)p^2
\end{equation}
With p given above, and $a = 0$ this equation has the solution
\begin{equation}
q= \frac{+\sqrt{e^2} \pm 2R_g/r \sqrt{e^2-(1-2R_g/r)}}{1-2R_g/r}
\end{equation}
where the positive sign on the first radical has been taken to assure that
time proceeds in a positive direction during the fall, and a positive
second term again corresponds to outgoing coordinates.
Since $v^r=p/q$, it is a straightforward matter to substitute for $v^r$ in the
original metric equation and examine the limit as $R_g/r \rightarrow 1/2$.
In either ingoing or outgoing coordinates, we find that $ds^2 \rightarrow 0$
as $R_g/r \rightarrow 1/2$. Thus the radial free-fall 
geodesics become null upon reaching the horizon.

It is of interest, however, that in the outgoing coordinates ($+4R_g/r$)
as $R_g/r \rightarrow 1/2$ one finds
$v^r \rightarrow 0, q \rightarrow \infty, p \rightarrow -e$. Thus it takes
an infinite coordinate time, but only a finite proper time to cross the horizon,
which is the same as the well-known Schwarzschild result.
In the ingoing coordinates, one obtains $v^r \rightarrow -c,
q \rightarrow e, p \rightarrow -e$. In this case, only a finite coordinate time
would be required to cross the horizon and coordinate speed,
$v^r$ does not change sign there. In
effect, one can continue calculations through the horizon
without reversing either the roles or
directions of change of r and t, but at the expense of ignoring that $ds^2 =0$
violates the SPOE requirement for timelike world line completeness.
In either case, it is interesting to observe that the
physical three-speed approaches that of light at the horizon
(Landau \& Lifshitz 1975).
\begin{equation}
V^2 = (\frac{dl}{d\tau_s})^2 = c^2\frac{({g_{0r}g_{0r}} - g_{rr}g_{00})v^r v^r}
        {(g_{00} + g_{0r} v^r)^2}
\end{equation}
Here we find $V \rightarrow c$ as $g_{00} \rightarrow 0$.
Both the null geodesic and $V \rightarrow c$ are consequences of the vanishing
of $g_{00}$ rather than arising from a singular metric. Few would
argue that the event horizon corresponding to the vanishing of $g_{00}$
is not a surface of infinite redshift, which is
what produces the null geodesic result, $d\tau = dt/(1+z) \rightarrow 0$.
Finally, it should be mentioned that the vanishing of $g_{00}$ for
$r > 0$ is actually a result of a failure to apply appropriate boundary
conditions for the solutions of the Einstein equations for a point mass
(Abrams 1979, 1989).

For suitably small energy, bound orbits occur. Turning
points for which $dr/d\tau = 0$ can be found by examining the
effective potential, which consists of all terms to the
right of $e^2$ in Equation 2. At minima of
the effective potential we find circular
orbits for which
\begin{equation}
a^{2}=\frac{1}{R_g/r-3 (R_g/r)^{2}}
\end{equation}
$R_g/r=1/3$ holds at the location of an
unstable circular orbit for photons (see below).
From which we see that if $p_\phi$ is non-zero there are no trajectories
for particles with both mass and angular momentum that exit from within $R_g/r=1/3$.
Thus particles with both mass and angular momentum
can't escape from within the photon sphere. The minimum energy
required for a circular orbit would be.
\begin{equation}
E=m_{0}c^{2}\frac{(1-2 R_g/r)}{\sqrt{(1-3 R_g/r)}}
\end{equation}
In fact, however, there is an innermost marginally stable orbit for
which the first two derivatives with respect to 1/r of the
effective potential vanish. This has no Newtonian physics counterpart,
and yields the well-known results: $R_g/r=1/6$, $a^2 = 12$ and $e^2 =8/9$
for the marginally stable orbit of radius $r_{ms} = 6GM/c^2$.

For a particle beginning a spiral descent from $r_{ms}$ with $e=\sqrt{8/9}$,
there follows:
\begin{equation}
(\frac{dr}{d\tau})^2=c^2\frac{(6R_g/r-1)^3}{9}
\end{equation}
If observed by a stationary observer located at coordinate r, it
would be observed to move with radial speed
\begin{equation}
V_r = \frac{\sqrt{2}c(6R_g/r-1)^{3/2}}{4}.
\end{equation}
Again, $V_r$ approaches c as $R_g/r$ approaches 1/2.
A distant observer would would find the angular frequency of the spiral motion
to be
\begin{equation}
\frac{1}{2\pi}\frac{d\phi}{dt} = \sqrt{9\times 12/8}(c^3/GM)(R_g/r)^2(1-2R_g/r)/2\pi
\sim 1.18\times 10^5 (R_g/r)^2(1-2R_g/r)/m ~~~Hz
\end{equation}
For a 10 M$_\odot$ GBHC ($m = 10$), this has a maximum of 437 Hz and
some interesting possibilities for generating many QPO frequencies, both
high and low. For red shifts such that $R_g/r \approx 1/2$, the spiral
frequency is  $2950/(1+z)^2$ Hz.\\
\\
{\bf Photon Trajectories:} \\
The energy-momentum equation for a particle with $m_0=0$ can be rearranged as:
\begin{equation}
(1-2R_g/r)^2(\frac{p_rGM}{p_\phi c^2})^2 = (\frac{d(R_g/r)}{d\phi})^2 =
(\frac{GME}{p_\phi c^3})^2 - (R_g/r)^2(1-2R_g/r)
\end{equation}
The right member has a maximum value of 1/27 for $R_g/r=1/3$. There is an
unstable orbit with $d(R_g/r)/d\phi = 0$ for $R_g/r=1/3$. To simply have
$d(R_g/r)/d\phi$ be real requires $p_\phi c^3/GME < \sqrt{27}$. But
$E=(1+z)p c$, where p is the entire momentum of the photon, and
$1+z = (1-2R_g/r)^{-1/2}$ its red
shift if it escapes to be observed at a large distance. Its azimuthal
momentum component will be $p_\phi/r$. Thus its escape cone is defined by:
\begin{equation}
(\frac{p_\phi}{r p})^2 < 27(R_g/r)^2(1-2R_g/r)
\end{equation}

\appendix{\bf{B. Pair Plasma Photosphere Conditions}}\\
Although we used a characteristic temperature of a pair plasma
to locate the photosphere and find its temperature, essentially the
same results can be obtained in a more conventional way.
The photosphere condition is that (Kippenhahn \& Weigert 1990):
\begin{equation}
n\sigma_T l = 2/3,
\end{equation}
where $n$ is the combined number density of electrons and positrons
in equilibrium with a photon gas at temperature T,
$\sigma_T = 6.65\times 10^{-25} cm^2$ is the Thompson scattering
cross section and $l$ is a
proper length over which the pair plasma makes the transition from
opaque to transparent. Landau \& Lifshitz (1958) show that
\begin{equation}
n=\frac{8\pi}{h^3}\int_{0}^{\infty} \frac{p^2dp}{\exp{(E/kT)}+1}
\end{equation}
where p is the momentum of a particle, $E=\sqrt{p^2c^2+m_e^2c^4}$,
k is Boltzmann's constant, $h$ is Planck's constant and $m_e$, the mass of an
electron. For low temperatures such that $kT < m_ec^2$ this becomes:
\begin{equation}
n \approx 2(\frac{2 \pi m_ekT}{h^2})^{3/2} \exp{(-m_ec^2/kT)}
\end{equation}
It must be considered that the red shift may change significantly over
the length $l$, and that $(1 + z_p)$ will likely be
orders of magnitude smaller than $(1 +z_s)$. Neglecting algebraic signs,
we can differentiate Equation (26) to obtain the coordinate length over
which z changes significantly as:
\begin{equation}
\delta r = \frac{R_g\delta z}{(R_g/r)^2(1+z)^3}
    \approx \frac{R_g}{(R_g/r)^2(1+z)^2}
\end{equation}
where we have taken $\delta z \approx (1+z)$. For values of z
appropriate here we take $R_g/r = 1/2$. We estimate
$l=\delta r (1+z) = 4R_g/(1+z)$ and replace $(1+z)$ with $T/T_\infty$.
Substituting expressions for $l$ and $n$ into the photosphere
condition and substituting for $T_\infty$ from equation 30
of the main text, equation 1 yields a transcendental equation for T.
For a GBHC with $z=10^8$ and $m=10$, its solution is $T_p = 3.3\times 10^8 K$
and $(1+z_p)=2500$. Then using the radiation pressure balance condition
in the pair atmosphere, we find $T_s^4 = T_p^4(1+z_s)/(1+z_p)$, from
which $T_s = 4.6\times 10^9 K$. The number density of particles
at the photosphere is $n=4\times 10^{20}$ and $10^9$ times larger at
the MECO surface. Nevertheless, the radiation pressure exceeds the
pair particle pressure there by ten fold. This justifies our
use of radiation dominated pressure in the pair atmosphere.
For an AGN with $1+z =10^8$ and $m=10^7$, we obtain photosphere and
surface temperatures of $2\times 10^8 K$ and $1.4\times 10^9 K$, respectively,
and $(1+z_p) = 50000$. We note that the steep temperature dependence of the pair
density would have allowed us to find the same photosphere temperature
within a few percent for any reasonable choice of $l$ from $10^3$ to $10^6$ cm.
In the present circumstance, we find $l=4R_g/(1+z) = 2.4\times 10^4$ cm.
This illustrates the extreme curvature of spacetime as the corresponding
coordinate interval thickness of the pair atmosphere
for the distant observer is only $\delta r \sim 10$ cm.\\

\appendix{\bf{C. Magnetosphere - Disk Interaction}}\\
We have found the inner accretion disk of MECO to be gas pressure dominated.
Impact pressure is not sufficient to push the magnetosphere inward. The
frequently used balance of impact and magnetic pressures to determine the
inner radius of the disk is not applicable. What is required instead is that
the stagnation pressure match the magnetic stress. Here we show that the same
inner radius scaling is obtained by merely requiring that the magnetosphere
remove disk angular momentum at an appropriate rate.
The torque per unit volume of plasma in the disk threaded by
magnetic field is given by
$r \frac{B_z}{4\pi} \frac{\partial B_{\phi}}{\partial z}
\sim r \frac{B_zB_{\phi}}{4\pi H}$, where H is the disk half thickness. Thus the
rate at which angular momentum would be removed from the disk would be
\begin{equation}
\dot{m}(v_K - 2\pi\nu_s r) = r\frac{B_zB_{\phi}}{4\pi H}(4\pi H \delta r).
\end{equation}
where $v_K$ is the Keplerian orbit speed and $\nu_s$ the spin frequency of
the central object.
The conventional expression for the magnetosphere radius can be obtained
with two additional assumptions: (i) that the field is fundamentally
a dipole field that is reshaped by the surface currents of the inner disk and
(ii) that $B_{\phi} = \lambda B_z (1 - 2\pi\nu_s r/v_k)$, where $\lambda$
is a dimensionless constant of order unity. This form
accounts for the obvious facts that $B_{\phi}$ should go to zero at
$r_c$, change sign there and grow in magnitude at greater distances from $r_c$.
In fact, however, we should note that
we are only describing an average $B_{\phi}$ here, because it is possible that
the field lines become overly stretched by the mismatch between
magnetospheric and Keplerian disk speeds, then break and reconnect across
the disk. This type of behavior leads to high frequency oscillations and
has been described in numerical simulations.
With these assumptions we obtain
\begin{equation}
r = (\frac{\lambda \delta r}{r})^{2/7}(\frac{\mu^4}{GM\dot{m}^2})^{1/7}
\end{equation}
In order to estimate $\delta r/r$,
we choose an object for which few would quibble about it being magnetic;
namely an atoll class NS. The rate of spin is typically 400 - 500 Hz,
the co-rotation radius is $\sim 26$ km, and the maximum luminosity
for the low state is $\sim 2\times 10^{36} = GM\dot{m}/2r$ erg/s,
from which $\dot{m} = 5.5\times 10^{16}$ g/s, for $M=1.4 M_\odot$.
For a magnetic moment of $\sim 10^{27}$ gauss cm$^3$, we find
that ($\frac{\lambda \delta r}{r})^{2/7} \sim 0.3$. Thus if $\lambda \sim 1$,
then $\delta r/r \sim 0.013$; i.e., the boundary region is suitably small,
though larger than the scale height of the trailing disk. In this small region
the flow changes from co-rotation with the magnetosphere to Keplerian. When
its inner radius is inside $r_c$, its weight is not entirely supported
by centrifugal forces and it provides the `dead-weight' against the
magnetopause.

\appendix{\bf{D. New Observations}}\\
The equations used in our previous work (Robertson \& Leiter 2002)
are repeated here for analysis of a few new observations.
Using units of  $10^{27}$
gauss cm$^3$ for magnetic moments, $100$ Hz for spin, $10^6$ cm for
radii, $10^{15}$ g/s for accretion rates, solar mass units,
$\lambda \delta r/r = 0.013$ and otherwise obvious notation, we found the
magnetosphere radius to be (Equation (2), Appendix C):
\begin{equation}
r_m~=~8\times 10^6 {(\frac{\mu_{27}^4}{m \dot{m}_{15}^2})}^{1/7}~~~~ cm
\end{equation}
A co-rotation radius of:
\begin{equation}
r_c = 7\times 10^6{(\frac{m}{\nu_2^2})^{1/3}} ~~~cm
\end{equation}
The low state luminosity at the co-rotation radius:
\begin{equation}
L_c = 1.5 \times 10^{34} \mu_{27}^2 {\nu_2}^3 m^{-1}~~~~erg/s
\end{equation}
High state luminosity for accretion reaching the central object:
\begin{equation}
L_s = \xi \dot{m}c^2 = 1.4\times 10^{36} \xi \mu_{27}^2 \nu_2^{7/3} m^{-5/3}~~~erg/s
\end{equation}
Where $\xi \sim 1$ for MECO and $\xi = 0.14$ for NS is the
efficiency of accretion to the central surface.
We calculate the quiescent luminosities in the soft x-ray band from
0.5 - 10 keV using the correlations of Possenti et al. (2001) with
spin-down energy loss rate as:
\begin{equation}
L_q = \beta \dot{E} = \beta 4 \pi^2 I \nu \dot{\nu}
\end{equation}
where $I$ is the moment of inertia of the star, $\nu$ its rate of spin and
$\beta$ a multiplier that can be determined from the new
$\dot{E} - L_q$ correlation for given $\dot{E}$;
i.e., known spin and magnetic moment. (In previous work we
had used $\beta = 10^{-3}$ for all objects.)
We assume that the luminosity is that of a spinning
magnetic dipole for which $\dot{E} = 32\pi^4 \mu^2 \nu^4/3c^3$,
(Bhattacharya \& Srinivasan 1995)
where $\mu$ is the magnetic moment.
Thus the quiescent x-ray luminosity would then be given by :
\begin{equation}
L_q = \beta \times \frac{32 \pi^4 \mu^2 \nu^4}{3c^3}
= 3.8 \times 10^{33} \beta \mu_{27}^2 \nu_2^4 ~~~~~erg/s
\end{equation}

As the magnetic moment, $\mu_{27}$, enters each of the luminosity equations
it can be eliminated from ratios of these luminosities, leaving relations
involving only masses and spins. For known masses, the ratios then
yield the spins. Alternatively, if the spin is known from
burst oscillations, pulses or spectral fit determinations of $r_c$,
one only needs one measured luminosity
to enable calculation of the remaining $\mu_{27}$ and $L_q$.
For most GBHC, we
found it to be necessary to estimate the co-rotation radius from
multicolor disk fits to the thermal component of low state spectra.
The reason for this is that the luminosities are seldom available across
the whole spectral hardening transition of GBHC.
For GBHC, it is a common finding that the low state inner disk radius is much
larger than that of the marginally stable orbit (e.g. Markoff, Falcke \&
Fender 2001, $\dot{Z}$ycki, Done \& Smith 1997a,b 1998, Done \& $\dot{Z}$ycki
1999, Wilson \& Done 2001). The presence of a magnetosphere
is an obvious explanation. Given an inner disk radius at the spectral
state transition, the GBHC spin frequency follows from the Kepler
relation $2 \pi \nu_s = \sqrt{GM/r^3}$.

{\bf Data:} \\
The third accreting millisecond pulsar, {\bf XTE J0929-314} has been found
(Galloway et al. 2002) with $\nu_s = 1/P = 185$ Hz and period derivative
$\dot{P} = 2.69\times 10^{-18}$, from which the magnetic field
(calculated as $3.2\times \sqrt(P\dot{P})$ is $3.9\times 10^9$ gauss.
This is typical of a Z source. Assuming a NS radius of 13 km,
the magnetic moment is $BR^3=8.5\times 10^{27}$ gauss cm$^3$.
The calculated low state limit co-rotation luminosity is $L_c=4.9\times 10^{36}$
erg/s. Approximately 40\% of this would be the luminosity in the (2 - 10 keV)
band. This yields an expected flux of $2\times 10^{-10}$ erg/cm$^2$/s for
a distance of 9 kpc. This corresponds to the knee of the published light curve
where the luminosity begins a rapid decline as the propeller becomes
active. Similar breaking behavior has been seen in Sax J1808.4-3659 and
GRO J1655-40 at propeller onset. The predicted 0.5-10 keV band luminosity
is $L_q = 1.3\times 10^{33}$ erg/s.

The second accreting millisecond pulsar {\bf XTE J1751-305} was found with a
spin of 435 Hz. (Markwardt et al. 2002) Its spectrum has been analyzed
(Miller et al. 2003). We find a hard state luminosity of
$3.5\times 10^{36}$ erg/s ($d=8$ kpc) at the start of the rapid
decline which is characteristic of the onset of the propeller
effect. We take this as an estimate of $L_c$.
From this we estimate a magnetic moment of $1.9\times 10^{27}$
gauss cm$^3$ and a quiescent luminosity of $5\times 10^{33}$ erg/s. An
upper limit on quiescent luminonosity of $1.8\times 10^{34}$ erg/s
can be set by the detections of the source in late April 2002, as
reported by Markwardt et al. (2002).

The accreting x-ray pulsar, {\bf GRO J1744-28} has long been cited for exhibiting
a propeller effect. Cui (1997) has given its spin frequency as 2.14 Hz and
a low state limit luminosity as $L_c = 1.8\times 10^{37}$ erg/s (2 - 60 keV.),
for a distance of 8 kpc. These imply a
magnetic moment of $1.3\times 10^{31}$ gauss cm$^3$ and a
magnetic field of $B=5.9\times 10^{12}$ gauss for a 13 km radius.
It spin-down energy loss rate should be $\dot{E} = 1.4\times 10^{35}$ erg/s
and its quiescent luminosity, $L_q = 3\times 10^{31}$ erg/s.
Due to its slow spin, GRO J1744-28 has a large co-rotation radius of
280 km. A mass accretion rate of $\dot{m}= 5.4\times 10^{18}$ g/s is needed
to reach $L_c$. Larger accretion rates are needed to reach the star surface,
but such rates distributed over the surface would produce luminosity in
excess of the Eddington limit. The fact that the magnetic field is strong
enough to funnel a super-Eddington flow to the poles is the likely reason for the
type II bursting behavior sometimes seen for this source. In addition to
its historical illustration of a propeller effect, this source exemplifies
the inverse correlation of spin and magnetic field strength in accreting
sources. It requires a weak field to let an accretion disk
get close enough to spin up the central object. For this reason we
expect Z sources with their stronger B fields to generally spin more
slowly than atolls.

The accreting pulsar, {\bf 4U0115+63}, with a spin of 0.276 Hz and a magnetic
field, derived from its period derivative, of $1.3\times 10^{12}$ gauss
(yielding $\mu = 2.9\times 10^{30}$ gauss cm$^3$ for a 13 km radius)
has been shown (Campana et al. 2002) to exhibit a magnetic propeller
effect with a huge luminosity interval from $L_c = 1.8\times 10^{33}$
erg/s to $L_{min} = 9.6\times 10^{35}$ erg/s. $L_c$ held steady precisely at
the calculated level for a lengthy period before luminosity began increasing.
Due to the slow spin of this star, its quiescent luminosity, if
ever observed, will be just that emanating from the surface. Its spin-down
luminosity will be much too low to be observed.

The atoll source {\bf 4U1705-44} has been the subject of a recent study
(Barret \& Olive 2002) in which a Z track has been displayed in a
color-color diagram. Observations labeled as 01 and 06 mark the end
points of a spectral state transition for which the luminosity ratio
$L_{min}/L_c = 25.6\times 10^{36}/6.9\times 10^{36} = 3.7$
can be found from their Table 2. These yield $\nu = 470$ Hz and
a magnetic moment of $\mu = 2.5\times 10^{27}$ gauss cm$^3$.
The spin-down energy loss rate is $1.2\times 10^{37}$ erg/s and
the 0.5 - 10 keV quiescent luminosity is estimated to be about
$5\times 10^{33}$ erg/s. At the apex of the Z track
(observation 12), the luminosity was $2.4\times 10^{37}$ erg/s (for
a distance of 7.4 kpc.); i.e., essentially the same as $L_{min}$.
Although 4U1705-44 has long been classified as an atoll source, it
is not surprising that it displayed the Z track in this outburst as
its 0.1 - 200 keV luminosity reached 50\% of the Eddington limit.

Considerable attention was paid to reports of a truncated accretion
disk for the GBHC, {\bf XTE J1118+480} (McClintock et al 2001) because
of the extreme interest in advective accretion flow (ADAF) models
for GBHC (Narayan, Garcia \& McClintock 1997). McClintock et al, fit
the low state spectrum to a disk blackbody plus power law model and found
that the disk inner radius would be about 35R$_{schw}$, or 720 km for
7 M$_\odot$. Using this as an estimate of the co-rotation radius we
find the spin to be 8 Hz. The corresponding low state
luminosity of $1.2\times 10^{36}$ erg/s (for $d= 1.8$ kpc) lets us
find a magnetic moment of $10^{30}$ gauss cm$^3$. The calculated spin-down
energy loss rate is $1.5\times 10^{35}$ erg/s and the quiescent luminosity
would be about $3\times 10^{31}$ erg/s.

A rare transition to the hard state for {\bf LMC X-3} (Soria, Page \& Wu 2002,
Boyd et al. 2000) yields
an estimate of the mean low state luminosity of $L_c = 7\times 10^{36}$
erg/s and the high state luminosity in the same 2 - 10 keV band is
approximately $6\times 10^{38}$ erg/s at the end of the transition to the
soft state. Taking these as $L_c$ and $L_{min}$ permits the estimates of
spin $\nu = 16$ Hz and magnetic moment $\mu = 8.6\times 10^{29}$ gauss cm$^3$,
assuming 7 M$_\odot$. From these we calculate a quiescent luminosity of
$10^{33}$ erg/s.


\begin{figure}
\plotone{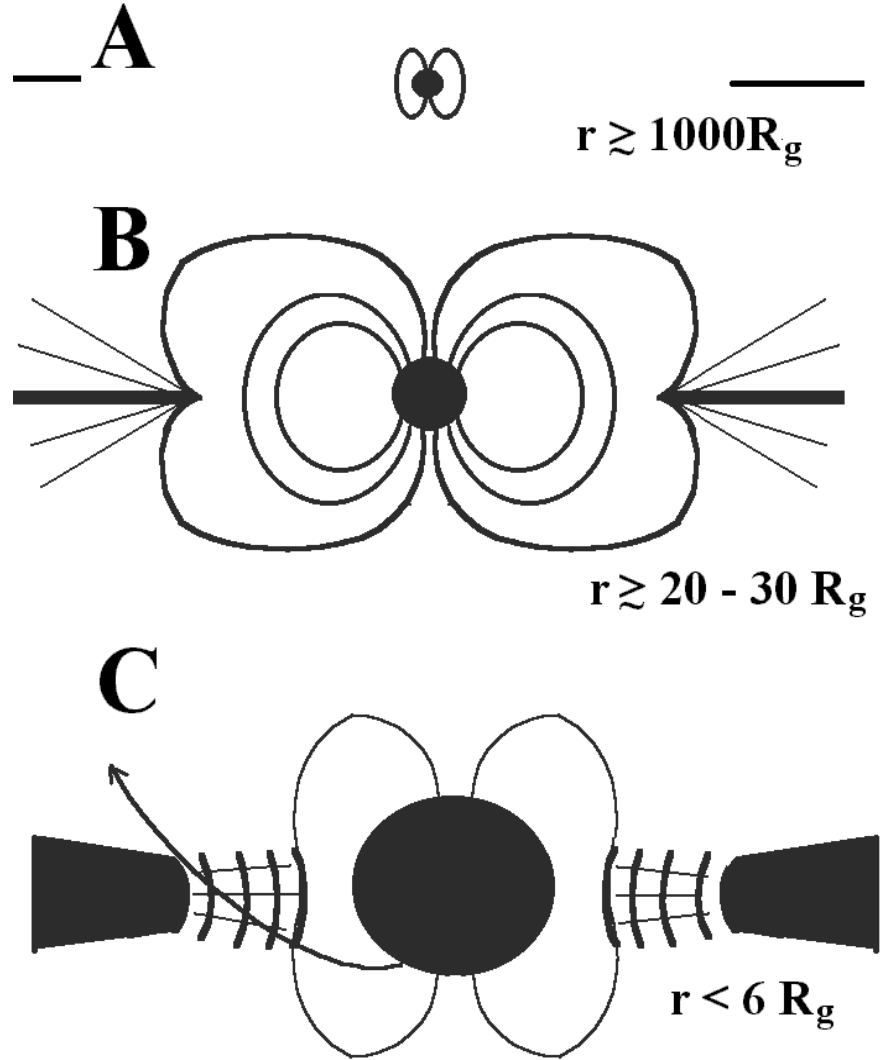}
\figcaption{MECO Spectral States:
{\bf A quiescent:}
Inner disk ablated, low accretion
rate to inner ablation radius $\sim 10^9 - 10^{10} cm$
generates optical emissions. Magnetic dipole radiation
drives hard power-law x-ray spectrum. Cooling NS surface x-ray
or quiescent MECO EUV emissions may be visible.
{\bf B. Low state:}
Thin, gas pressure dominated inner disk has large magnetically
dominated viscosity. The inner disk radius lies
between the light cylinder and co-rotation radii.
Disk winds and jets are driven by
the magnetic propeller. A hard spectrum is produced as most soft
x-ray photons from the disk are Comptonized by either outflow or corona.
Outflows of electrons on open magnetic field lines produce synchrotron
radiation.
Most of the outer disk is shielded from the magnetic field of the central
object as surface currents in the inner disk change the topology of the
magnetopause.
{\bf C: High state:}
Once the inner disk is
inside the co-rotation radius, the outflow and synchrotron emissions
subside. Relaxation oscillations may occur as radiation from the
central object momentarily drives the inner disk back outside the
co-rotation radius. A boundary layer of material beginning to
co-rotate with the magnetosphere may push the magnetopause to the star
surface for NS or inside $r_{ms}$ for MECO, where
a supersonic flow plunges inward until radiation
pressure stabilizes the magnetopause or interchange instabilities break
up the flow.
The MECO photosphere radiates a bright `ultrasoft'
thermal component. Bulk comptonization of many photons on spiral
trajectories crossing the disk produces a hard x-ray spectral tail.}
\end{figure}

\end{document}